\documentclass[prb,showpacs,amsmath,amssymb,twocolumn,letterpaper]{revtex4}

\usepackage{amssymb}
\usepackage{graphicx}
\usepackage{color}
\usepackage{dcolumn}
\usepackage{bm}
\usepackage{subfigure}
\usepackage{natbib}

\begin{document}

\newcommand{\tc}{$T_{c}$ }
\newcommand{\ts}{$T^{*}$ }
\newcommand{\tn}{$T_{N,Ce}$ }
\newcommand{\tk}{$T_{K}$ }
\newcommand{\tco}{$T_{coh}$ }

\title{Search for pressure induced superconductivity in CeFeAsO and CeFePO iron pnictides}

\author{D. A. Zocco,$^{1}$ R. E. Baumbach,$^{1}$ J. J. Hamlin,$^{1}$ M. Janoschek,$^{1}$ I. K. Lum,$^{1}$ M. A. McGuire,$^{2}$ A. S. Sefat,$^{2}$ B. C. Sales,$^{2}$ R. Jin,$^{2}$ D. Mandrus,$^{2}$ J. R. Jeffries,$^{3}$ S. T. Weir,$^{3}$ Y. K. Vohra,$^{4}$ and M. B. Maple$^{1}$}
\affiliation{$^{1}$Department of Physics, University of California, San Diego, La Jolla, California 92093, USA\\
$^{2}$Materials Science \& Technology Division, Oak Ridge National Laboratory, Oak Ridge, Tennessee 37831, USA\\
$^{3}$Lawrence Livermore National Laboratory, Livermore, California 94551, USA\\
$^{4}$Department of Physics, University of Alabama at Birmingham, Birmingham, Alabama 35294, USA}

\date{\today}

\begin{abstract}
The CeFeAsO and CeFePO iron pnictide compounds were studied via electrical transport measurements under high-pressure. In CeFeAsO polycrystals, the magnetic phases involving the Fe and Ce ions coexist for hydrostatically applied pressures up to 15 GPa, and with no signs of pressure-induced superconductivity up to 50 GPa for the less hydrostatic pressure techniques. For the CeFePO single crystals, pressure further stabilizes the Kondo screening of the Ce 4$f$-electron magnetic moments.
\end{abstract}

\pacs{74.62.Fj, 75.20.Hr, 74.70.Dd}

\maketitle

Since the discovery of the iron-based superconductors more than two years ago,\cite{kamihara_2006_1, watanabe_2007_1, kamihara_2008_1} it has been shown that chemical substitution and the application of external pressure constitute two complimentary ways of tuning the lattice structure of these materials and commensurately affecting the values of the critical temperatures at which superconducting, magnetic and structural phases emerge. The replacement of La in LaFeAsO$_{1-x}$F$_{x}$ by heavier and smaller lanthanides enhances the superconducting transition temperature $T_{c}$, $e.$ $g.$, from 26 K to 55 K in SmFeAsO$_{1-x}$F$_{x}$.\cite{ren_2008_1} Likewise, the application of pressure increases \tc from 26 K to 43 K in the optimally doped LaFeAsO$_{1-x}$F$_{x}$,\cite{takahashi_2008_1, zocco_2008_1} and from 7 K to 14 K in LaFePO.\cite{hamlin_2008_1} On the other hand, superconductivity (SC) has only been found to be induced by the application of external pressure in the LaFeAsO parent compound.\cite{okada_2008_1} The relatively limited amount of experimental work on the $Ln$Fe$Pn$O ($Ln$ = lanthanide, $Pn$ = As, P) parent compounds under pressure \cite{hamlin_2008_1, okada_2008_1, igawa_2009_1, takahashi_2009_1, chu_2009_1, kawakami_2009_1, okada_2010_1} motivated us to extend our effort to two of the members of the ``1111'' family, CeFeAsO and CeFePO.

The compound CeFeAsO is a semimetal, with coexisting Fe-3$d$ and Ce-4$f$ antiferromagnetic (AFM) phases below 4 K, and becomes a superconductor below 41 K when 16\% of oxygen is substituted by fluorine.\cite{chen_2008_1} It was found that CeFeAsO undergoes a lattice distortion from a tetragonal to an orthorhombic structure near 155 K, followed by the formation of a commensurate spin density wave (SDW) on the Fe sublattice below 140 K,\cite{zhao_2008_1} although it has been recently found that the temperature difference between the structural and the magnetic transitions of CeFeAsO is strongly sample dependent.\cite{jesche_2010_1} On the other hand, CeFePO is a heavy fermion metal ($\gamma \sim$ 700 mJ/molK$^{2}$) with no magnetic order found to date.\cite{bruning_2008_1} Recently,\cite{luo_2010_1, delacruz_2010_1} the isoelectronic substitution of As by P has revealed a rich phase diagram, in which an antiferromagnetic (AFM) and a ferromagnetic (FM) quantum critical point (QCP) are found at intermediate compositions. The experiments reported herein were carried out, in part, to address whether the Ce ion valences in CeFeAsO and CeFePO are sensitive enough to pressure to increase the electron concentration of the Fe-As/P layers by driving Ce from the 3+ towards the 4+ valence state. For CeFeAsO, this could have the effect of decreasing the fraction of the Fermi surface (FS) gapped by the long-range Fe magnetism, resulting in a dependence of the transition temperatures on pressure similar to that observed for F substitution. Measurements made under nearly hydrostatic conditions in a piston-cylinder clamp indicate that the Fe-SDW ordering temperature \ts decreases with pressure. In contrast, the N\'{e}el temperature \tn for the AFM ordering of Ce ions is observed to increase with pressure. For the CeFePO system, we found that pressure stabilizes the Kondo screening of the Ce 4$f$-electron magnetic moments.

Polycrystalline samples of CeFeAsO were prepared as described in Ref.~\onlinecite{mcguire_2009_1}. For the CeFePO single crystals, the molten Sn:P flux technique described in Ref.~\onlinecite{baumbach_2009_1} was used. Electrical resistivity measurements were performed for 1.2 K $\leq$ $T$ $\leq$ 300 K under nearly-hydrostatic pressure conditions up to 2.7 GPa employing a beryllium-copper piston-cylinder clamped cell, using a 1:1 mixture by volume of $n$-pentane and isoamyl alcohol as the pressure medium and a Teflon capsule. In order to attain higher pressures, a Bridgman-anvil clamped cell and a mechanically loaded ``designer'' diamond-anvil cell (DAC) were used, as described in more detail in Ref.~\onlinecite{zocco_2008_1}, reaching maximum pressures 14.5 and 50 GPa, respectively, for each technique. For the Bridgman-anvil cell, we used 4 mm diameter anvil flats and quasi-hydrostatic solid steatite as pressure medium. For the DAC experiments, the gasket was made from a 200 $\mu$m thick MP35N foil pre-indented to 40-50 $\mu$m and a 100 $\mu$m diameter hole was drilled through the gasket using an electrical discharge machine (EDM). In this case, the sample was packed in the gasket hole, without pressure transmitting medium. Pressure gradients were inferred from the width of the superconducting transition of a Sn manometer for the hydrostatic ($\delta$$P$ $<$ 2\%) and Bridgman (2\% $<$ $\delta$$P$ $<$ 10\%) cells. For the DAC, the pressure was adjusted and determined at room temperature, using the fluorescence spectrum of a $\sim$ 10 $\mu$m chip of ruby located within the sample chamber but slightly offset from the center, where the voltage leads are located, with a $\delta$$P$ $\leq$ 15\%, inferred from the full width at half maximum (FWHM) of the fluorescence line. A standard four-lead technique and a Linear Research Inc. LR-700 AC resistance bridge with excitations smaller than 1 mA were used to measure the electrical resistance.

\begin{figure}
{\includegraphics[width=3.2in]{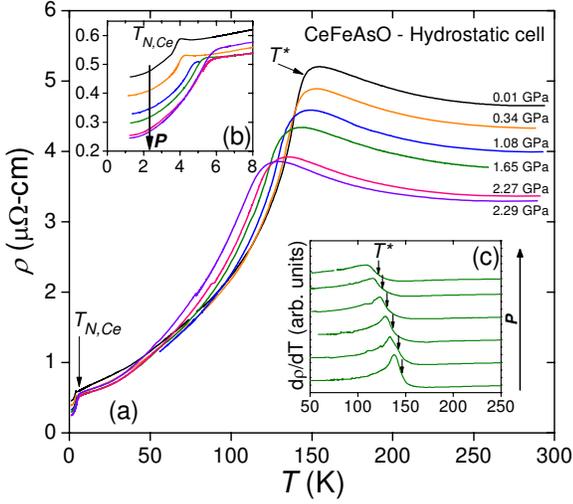}} \caption{(Color online) (a) Electrical resistivity $\rho$ versus temperature $T$ at various pressures $P$ for CeFeAsO polycrystalline samples, obtained in the hydrostatic clamped cell experiment. (b) Evolution of \tn at low temperatures. (c) d$\rho(T)$/d$T$ vs. $T$ curves for the hydrostatic experiment. The arrows indicate the position of $T^{*}$.} \label{CeFeAsO_1}
\end{figure}

\begin{figure}
{\includegraphics[width=3.21in]{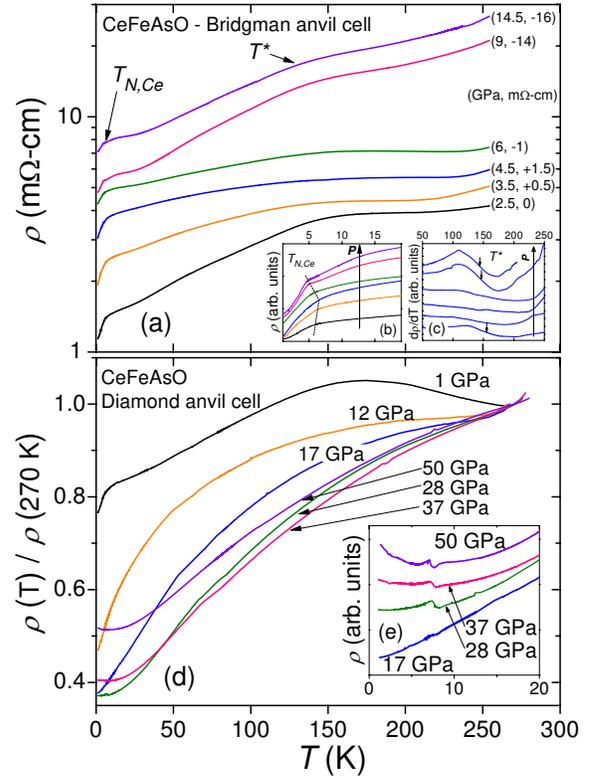}} \caption{(Color online) (a) Electrical resistivity $\rho$ versus temperature $T$ at various pressures $P$ for CeFeAsO polycrystalline samples for the Bridgman anvil cell experiment. The curves are shifted for clarity. The pressure and magnitude of the offset are indicated in parentheses at the right of the curves. (b) Low temperature detail, showing the evolution of \tn at various pressures. (c) d$\rho(T)$/d$T$ vs. $T$ curves for the Bridgman anvil cell experiment. (d) Diamond anvil cell resistivity curves normalized to 270 K. (e) Low temperature detail for the higher pressures.} \label{CeFeAsO_2}
\end{figure}

The electrical resistivity of polycrystalline CeFeAsO as a function of temperature, obtained in the hydrostatic experiment, is plotted in Fig.~\ref{CeFeAsO_1}. Two main features are identified: the peak-like feature centered at \ts $\sim$ 150 K, which was previously identified as the onset to SDW order of the Fe electronic spins, and the kink at \tn $\sim$ 4 K, attributed to the antiferromagnetism of the localized spin states of the Ce atoms.\cite{chen_2008_1} We defined \ts as the mid-point temperature in the derivatives d$\rho(T)$/d$T$, indicated with the vertical arrows in Fig.~\ref{CeFeAsO_1}c. For measurements using the hydrostatic clamp (Fig.~\ref{CeFeAsO_1}a), \ts decreases with applied pressure at a rate of $-$9 K/GPa, which is similar to the value obtained for LaFeAsO.\cite{okada_2008_1} On the other hand, \tn increases at a rate of +0.9 K/GPa, as shown in detail in Fig.~\ref{CeFeAsO_1}b. These results suggest that, with increasing pressure, more Fe $d$-electrons participate in the indirect interaction between the magnetic Ce ions. The increase of \tn with pressure can be qualitatively explained in both an SDW and a local moment picture for the AFM ordering of the Ce ions. In the SDW picture, the decrease with pressure of the FS fraction that is gapped by the Fe SDW results in an increase of the FS fraction that is available to be gapped by the Ce AFM state, thereby increasing \tn with pressure. In the local moment picture, the increase of the FS fraction that remains ``ungapped'' by the Fe SDW results in an increase of the density of states at the Fermi level, $N(E_{F})$. In addition, the increase of pressure results in an increase in the magnitude of the exchange interaction parameter $|J|$, which is well documented for the case of Ce ions in dilute alloys ($e. g.$, La$_{1-x}$Ce$_{x}$) or compounds ($e. g.$, SmS).\cite{schilling_1979_1,maple_2005_1} The increase of $|J|$ with pressure results from the increase of hybridization which can be visualized within the context of the Friedel-Anderson model through the Schrieffer-Wolff transformation $J \sim - \frac{\langle V_{fc}^{2}\rangle}{\varepsilon_{f}}$, where $\langle V_{fc} \rangle$ is the matrix element that admixes the Ce 4$f$-electron states and the conduction electron states, $\varepsilon_{f}=E_F - E_f$ is the $f$-electron binding energy, and $E_f$ is the centroid of the $f$-state, where the width $\Delta$ of the localized state is $\Delta\sim \pi N(E_{F})\langle V_{fc}^{2}\rangle$. A more complete picture, which includes the orbital angular momentum is given by the Coqblin-Schrieffer transformation (see, for example, Ref.~\onlinecite{maple_2005_1}). This situation should also produce Kondo lattice physics in which a nonmagnetic ground state forms below the Kondo temperature $T_{K}\sim T_{F}exp(-1/N(E_{F})|J|)$, where $T_{F}$ is the Fermi temperature. Since the N\'{e}el temperature \tn is expected to vary as \tn $\sim N(E_{F})J^{2}$, the existence of AFM ordering and the increase of \tn with pressure indicates that \tn $>$ $T_{K}$.\cite{doniach_1977_1}

\begin{figure}
{\includegraphics[width=3.4in]{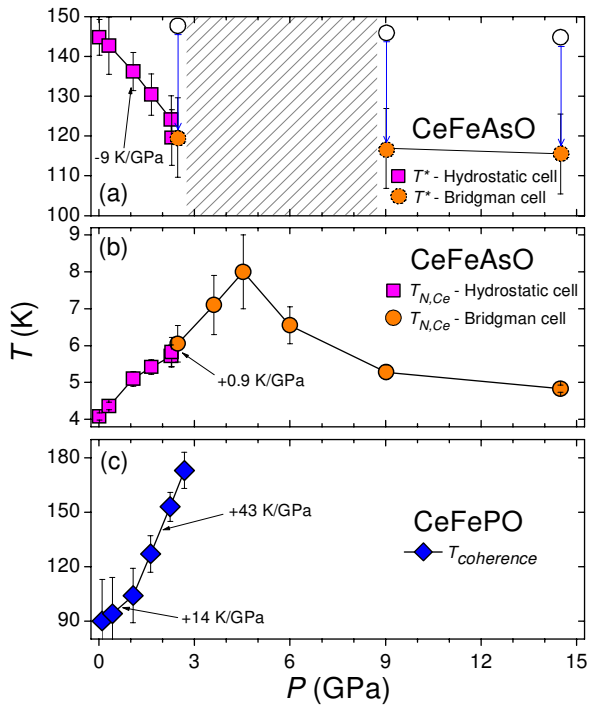}} \caption{(Color online) $T$-$P$ phase diagram for CeFeAsO (panels $a$ and $b$) and CeFePO (panel $c$). In $a$), data corresponding to the Bridgman-anvil technique are shown shifted to match the hydrostatic cell measurements. In the upper panel, the shaded region indicates the pressure range where the value of \ts is more difficult to determine from our measurements.} \label{phasediagram}
\end{figure}

Recently,\cite{pourovskii_2008_1} LDA+DMFT calculations showed that, for CeFeAsO, the hybridization potential $V_{fc}$ increases substantially with pressure, which is reflected as an exponential increase of $T_{K}$, from 10$^{-4}$ K at ambient pressure, to 100 K at 15 GPa. A rough comparison of our rate of suppression of \ts with that observed when oxygen is replaced by fluorine in the same compound \cite{chen_2008_1} with the subsequent appearance of superconductivity, motivated us to perform additional measurements on CeFeAsO in the higher pressure region. The Bridgman-anvil cell data are shown in Fig.~\ref{CeFeAsO_2}a, for 2.5 GPa $\leq$ $P$ $\leq$ 14.5 GPa. At the lowest pressure, the feature corresponding to \ts appears in the vicinity of 150 K, and then broadens at higher pressures. At 4.5 GPa, this feature is no longer distinguishable, as can also be seen in the very flat derivative curve in Fig.~\ref{CeFeAsO_2}c. At higher pressures, the broad feature re-emerges and persists up to 14.5 GPa, although its association with the SDW can no longer be verified. An opposite trend is observed for the antiferromagnetic ordering of the Ce moments (Fig.~\ref{CeFeAsO_2}b). \tn increases from 6 K to a maximum value of 8 K at the same rate as in the hydrostatic cell experiments, and at 4.5 GPa, \tn begins to decrease and reaches 5 K at 14.5 GPa. The hydrostatic cell and Bridgman anvil cell results are summarized in the $T$-$P$ phase diagram of Fig.~\ref{phasediagram}. Since the values of \ts at 2.5 GPa do not match for the hydrostatic and Bridgman-anvil techniques, presumably due to sample dependence or larger strains in the Bridgman cell technique, but the slopes d$T^*$/d$P$ are similar, we plotted the Bridgman-anvil experiment values of \ts offset down to match the last point obtained in the hydrostatic experiment, in order to facilitate comparison. The shaded area in Fig.~\ref{phasediagram}a indicates the pressure region in which \ts is more difficult to identify from our measurements.

\begin{figure}
{\includegraphics[height=2.6in]{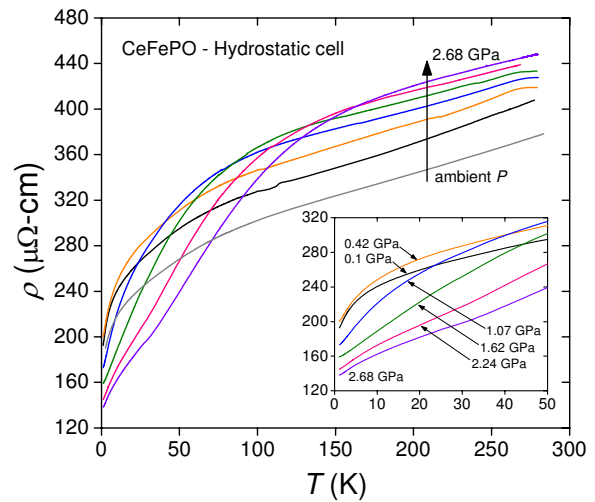}} \caption{(Color online) Electrical resistivity $\rho$ versus temperature $T$ of CeFePO single crystals at pressures $P$ of 0.01, 0.10, 0.42, 1.07, 1.62, 2.24 and 2.68 GPa, obtained with the hydrostatic clamped cell technique. $Inset$: low temperature region.} \label{CeFePO}
\end{figure}

Fig.~\ref{CeFeAsO_2}d shows the diamond anvil cell (DAC) electrical resistivity vs. temperature for CeFeAsO for 1 GPa $\leq$ $P$ $\leq$ 50 GPa (curves were normalized to the corresponding values at 270 K). At the lowest pressure (1 GPa), both the Fe-SDW and the Ce-AFM features were observed. At higher pressures, the Fe-SDW shoulder smeared out, and even the sharper Ce-AFM transition disappeared and was replaced by minima and upturns at the highest pressures (see Fig.~\ref{CeFeAsO_2}e).

It is worth emphasizing the lack of pressure induced superconductivity (SC) in the CeFeAsO samples. So far, SC in an undoped ``1111'' iron arsenide was reported exclusively in LaFeAsO,\cite{okada_2008_1} where only one electrical resistivity curve dropped completely to zero below $T_c$. Also, in that work, a 1:1 mixture of Fluorinert FC70/77 was utilized as the pressure transmitting medium which is known to be non-hydrostatic above 0.8 GPa.\cite{sidorov_2005_1} A detailed discussion of the effects of non-hydrostatic pressure media in the ``1111'' and ``122'' Fe-pnictide families was recently reported by W. Bi \textit{et al.}.\cite{bi_2010_1} If we assume that the pressure induced SC in polycrystalline LaFeAsO is unrelated to oxygen deficiency, we might then speculate that the appearance of SC could be related to significant strain which can develop even in hydrostatic media, due to the anisotropic compressibility of adjacent randomly oriented grains. In polycrystalline CeFeAsO, however, SC didn't appear despite the very non-hydrostatic conditions obtained in the diamond anvil cell, although the response to strain may be opposite for the 122 materials compared to the 1111 materials. It would be interesting to see if pressure induced SC could appear under nearly strain free conditions where single crystalline $Ln$FeAsO samples were compressed under hydrostatic conditions.

We also note that in the iron-pnictide compounds, the shape of the Fe$Pn_{4}$ tetrahedra is highly correlated with the maximum \tc achieved.\cite{delacruz_2008_1, zhao_2008_1, lee_2008_1, mito_2008_1, kimber_2009_1, parker_2010_1} Theoretical work has shown that the Fermi surface topology is very sensitive to the Fe-$Pn$ bond length as well as the deviation of $Pn$-Fe-$Pn$ angle from the value of 109.47$^\circ$ corresponding to a regular tetrahedron, mainly due to the rather high degree of covalency of the Fe-$Pn$ bond.\cite{mcqueen_2008_1, vildosola_2008_1, wilson_2010_1} A comparison between the pressure and the doping dependences of the $Pn$-Fe-$Pn$ angle of LaFeAsO and CeFeAsO would, perhaps, help us to understand the appearance of the superconducting state in these materials.

Fig.~\ref{CeFePO} shows the electrical resistivity $\rho(T)$ for a single crystalline CeFePO sample for different pressures applied in the hydrostatic piston-cylinder clamp (ambient $\leq$ $P$ $\leq$ 2.7 GPa). The resistivity exhibits a typical Kondo lattice behavior throughout the entire pressure range, \textit{i. e.}, $\rho(T)$ decreases weakly at high temperatures and drops strongly below $T_{coh}$, the Kondo-coherence temperature. It is well known that for Ce compounds, pressure increases the value of the hybridization matrix element $V_{fc}$ between $4f$- and conduction electrons,\cite{maple_1973_1} causing an increase of $T_{coh}$. In the CeFePO case, as pressure increases, \tco shifts towards higher temperatures, with and initial slope of +14 K/GPa, and at a much higher rate of +43 K/GPa above 1 GPa (Fig.~\ref{phasediagram}c). A broad feature develops below 30 K, which is apparent at 2.68 GPa, as can be seen in the \textit{inset} of Fig.~\ref{CeFePO}, the physical origin of which remains unclear. A possible explanation could be found in a recent study on CeRhSn,\cite{slebarski_2002_1} where a similar feature was observed near 6 K in the $\rho(T)$ data. This was attributed to non-Fermi liquid behavior in the proximity to a quantum critical point, possibly originating from inhomogeneous magnetic ordering of the spin-glass-type, produced by atomic disorder. We are currently performing magnetization and specific heat experiments at ambient pressure to test different scenarios, and the results will be published elsewhere.

In summary, we have presented high-pressure electrical resistivity measurements on CeFeAsO polycrystals and CeFePO single crystals. At low pressures, the Fe and Ce magnetic phases of CeFeAsO seem to compete with each other, and no superconductivity was found for pressures up to 50 GPa, which indicates that the application of pressure might affect  the electronic properties of CeFeAsO in a significantly different manner compared to chemical substitution. For CeFePO single crystals, pressure stabilizes the Kondo screening of the Ce 4$f$-electron magnetic moments, which is reflected as an increase of $T_{coh}$ with pressure.

High-pressure research at UCSD was supported by the National Nuclear Security Administration (NNSA) under the Stewardship Science Academic Alliance program through the U.S. Department of Energy (DOE) grant number DE-FG52-06NA26205. Crystal growth at UCSD was supported by the US DOE grants DE-FG02-04ER46105 and DE-FG02-04ER46178. Research at ORNL is sponsored by the Division of Materials Sciences and Engineering. Research at LLNL was supported by NNSA-DOE grant DE-AC52-07NA27344. Research at UAB was supported by NNSA-DOE grant DE-FG52-06NA26168. MJ acknowledges support from the Alexander von Humboldt Foundation.

\end{document}